# Black Lives Matter discourse on US social media during COVID: polarised positions enacted in a new event

Gillian Bolsover
University of Leeds
g.bolsover@leeds.ac.uk

**Abstract**

*Black Lives Matter has been a major force for social change in the US since 2014, with social media playing a core role in the development and proliferation of the movement. The largest protests in US history occurred in late May and early June 2020, following the death of George Floyd at the hands of Minneapolis police. This incident reignited widespread support for the BLM movement. The protests were notable not only for their size but also that they occurred at a time the US was still struggling to control the spread of the COVID-19 pandemic, with more than 20,000 new cases per day. With protest conditions and police crowd control tactics exacerbating disease spread and with COVID disproportionately affecting minority populations, it was hypothesised that participation in and support for the protests would involve a balancing act between the risks of systemic racism and of disease spread. However, social media data suggest that this was not the case, with discussion of the BLM movement replacing discussion of COVID on US social media. Neither supporters or opposers of the BLM movement or protest action mentioned COVID as a factor. Framings of the movement by BLM supporters largely replicated those of earlier studies, with new frames emerging surrounding the opposition the movement has received from racism, police militarisation and President Donald Trump. Discourse evidenced worrying levels of polarisation, hate, incivility and conspiracy content and bore many similarities to previously studied COVID discourse. This suggests that George Floyd's death, as yet another example of an African American man killed by US police, was largely seen through established, polarised identity positions that made reactions to the incident and resulting protest largely a foregone conclusion, established and articulated without reference to the ongoing pandemic.*

**COVID and systemic racism: twin challenges**

In the summer of 2020, the US was rocked by two very different political issues associated with opposite forms of individual political response. On the one hand, the country faced an unprecedented challenge posed by COVID-19, a novel and highly infectious respiratory illness, that has, to date, infected more than five million Americans. Compared to stringent restriction measures to control the spread of the virus in other countries, the US has seen a slow and limited response. There has been much resistance to social and economic controls to limit the spread of the virus in the US (Bolsover, 2020). President Donald Trump has largely opposed controls and has spread misinformation about virus treatments and preventative measures (Bolsover & Tokitsu Tizon, 2020).

In the midst of this public health crisis, longstanding issues of systemic racism once again hit a crisis point. This came at the end of May in response to the killing of 46-year-old father of five George Floyd. Police officers responding to a call about someone using a counterfeited $20 note ended up kneeling on Floyd's neck for close to eight minutes as he lay face down on the street restrained by several police officers. He died at the scene.

The widely circulated videos of the incident documented the latest in a number of deaths of black Americans at the hands of the police, including Eric Garner who suffocated in a police chokehold after repeating "I can't breathe," 12-year-old Tamir Rice who was holding a toy gun when he was shot dead by police, 18-year-old Michael Brown who was unarmed and shot six times after he stole a box of cigars and Breonna Taylor who was shot dead in her home while police officers executed a "no knock" search warrant (no drugs were found on the premises). These deaths, among countless others, have mobilised Americans, under the Black Lives Matter movement, to protest the unequal treatment of black Americans at the hands of the police and justice system.

The Black Lives Matter movement was founded in 2013 in response to the acquittal of a white man who fatally shot 17-year-old Trayvon Martin in an affluent Florida gated-community. Social media has played a key role in the foundation and building of the Black Lives Matter movement (Ince et al., 2017), particularly as the documentation on cell phone video has been crucial in building visibility and credibility for cases of excessive use of force by police. In the days following Floyd's death on 25 May 2020, the US saw protests across the country numbering between 15 and 26 million people, potentially the largest in US history (Buchanan et al., 2020). However, these protests came at a time that the US was still trying to control the COVID pandemic, with 20,000 new cases a day (CDC, 2020).

This case raises important questions about how different risks are being balanced by individuals engaging in these protests. Was participation in these protests underpinned by an active calculation that the risks of systemic racism outweighed the risks of disease spread? Did participants consciously recognise and attempt to mitigate disease spread at the protests? Were COVID and BLM viewed as separate issues or was the intersectionality of risk for African Americans discussed as part of protest discourse? What role did social media play in the building of the movement under changed risk conditions?

**The BLM movement and social media**

Over the past decade, social media has been an important tool in the organisation and proliferation of social movements. This has been especially true for the BLM movement that grew out of a Facebook post and was conceived as a social media movement around the hashtag #BlackLivesMatter (Ince et al., 2017). Social media, and Twitter in particular, allowed the stories of these individuals and the narratives of the movement to be distributed across the country without relying on mainstream media (Freelon et al., 2016). Social media has been important in distributing both the documentation of cases of police violence against individual African Americans and the documentation of violent police reactions to the BLM protests, instigating expressions of disbelief among casual observers on social media (ibid). Social media has also been a major source of information about the movement for the wider public (Cox, 2017) and the affect, behaviour and language associated with the movement on social media can predict on the ground protest participation (Choudhury et al., 2016).

As a distributed social movement, however, a variety of individuals interact with and contribute to discourse about BLM on social media, including individuals who oppose the BLM movement, appropriating and deploying the movements phraseology for their own goals (Ince et al., 2017). For instance, the hashtags #AllLivesMatter and #BlueLivesMatter were used by those who opposed or criticised the BLM movement, used (respectively) to deny specific violence against black Americans and to suggest lack of attention to the importance of police safety by BLM supporters (Carney, 2016).

During the first eleven months of the BLM movement in 2014, research found that discourse on social media using the BLM hashtag fell into five groups: expressions of solidarity and affirmations of the goals of the movement; discussion of tactics and actions of the movement; reports, documentation and reactions to police violence; discussions of the protests in Ferguson, Missouri following the shooting of Michael Brown; and counter protest messages blocking the goals, tactics and ideology of the movement (Ince et al., 2017).

The protests and riots that emerged following the shooting of Brown in August 2014 shocked the country. Photos of military-grade weapons being used on unarmed civilians, the shooting of wooden pellets and tear gas, and the arrest of two journalists legally filming these actions sparked debates about the militarisation of the police and excessive use of force by police not just against individual black people but also against those protesting in these cases (Li, 2014).

These tactics were repeated in the 2020 protests following the death of George Floyd. However, the ongoing coronavirus threat posed a new complication in this case. Not only would protest conditions facilitate the spread of the virus but the use of tear gas and crowd control and containment tactics by the police would aid the spread and worsen the symptoms of the virus (Harvey, 2020). BLM activists and allies faced an unprecedented challenge of balancing the risks of the virus against the risk of systemic racism, complicated by the intersectionality of these risks that has left ethnic minority populations disproportionately affected by the virus as well as local control measures that target areas in which the virus is actively spreading.

**COVID, minorities and intersectional impacts**

The COVID pandemic, which has to date killed more than 170,000 Americans, has disproportionally impacted ethnic minorities. Although racial data on COVID impacts is still incomplete, analysis of available data in the US found African American deaths from COVID were almost twice as much as would be expected based on their share of the population (Godoy & Wood, 2020). Similarly, in the UK, mortality rates for black men and women are 1.4 to 2 times higher than for white groups and individuals with south Asian backgrounds are 20% more likely to die after being admitted to hospital than white individuals (Haque, 2020).

A report by Public Health England concluded that in the UK these differences were not due to biological causes but rather the effects of structural inequalities, marginalisation and racism, moderated through lower socioeconomic status, occupation, living conditions and poorer health (Public Health England, 2020). As such, both the police violence that underpins the BLM movement and the threat of COVID for minority populations share root causes that the BLM movement aims to address.

With this in mind, the 2020 BLM protests present a unique case to understand how individuals approach and balance different forms of risk and engage in complicated calculations about political responses to societal issues. It is also an important step in the evolution of the BLM movement, representing the largest and most widespread protests on the US taking place under conditions of increased risk for protestors and increased public resistance to large-scale gatherings due to disease transmission risk.

In analysing this case, we are particularly interested to understand how individuals are balancing different forms of risk in their thinking about the protest movement. Do they feel the risks of police racism outweigh the risks of COVID? Do the systemic root causes of the impacts of these issues on black communities strengthen the arguments of the movement? Will there be increased pushback against the protests due to the COVID transmission risks incurred by protest action? Do the frames used by individuals in discussing the movement match those found in previous research or has the issue evolved in light of the current pandemic? Given the core role that social media has played in the BLM movement and in political discussions more widely in the US, we focus this analysis on online discourse on Twitter.

**Examining Twitter discourse**

Many research projects intending to monitor social media discourse on Twitter collect data from within a group of pre-selected hashtags and keywords. However, this strategy risks missing emergent or unselected topics. It is thus severely limited in its ability to speak to the body of online discourse, particularly during fast-moving events, and is subject to significant researcher bias based on the selection of hashtags and keywords to follow.

To avoid this limitation, this project collected a sample of data from all trending topics within the US between 25 May and 31 May, 2020. This period therefore captures the initial reaction, discussion and organisation of the protests, which occurred in the days following Floyd's death on 25 May.

Using custom Python scripts to interface with the Twitter API, the project collected the most recent 100 tweets associated with each of the top 50 trending topics in the 64 locations for which Twitter collates trends (including one for the entire country) every 15 minutes during the target period. This data collection captured 3,434 unique trending topics across the week-long period.

We then undertook a content analysis of each hashtag or keyword that had trended at least 50 times across the 64 locations. This means, for example, a coded topic could have been trending in almost all US locations in one 15-minute period or in just one US location for more than half a day. The content analysis employed an established coding frame used in previous research to determine the broad content of social media posts in a variety of political contexts and events (Bolsover, 2017, 2018). A total of 324 hashtags or keywords trended more than 50 times across the 64 locations during the period.

Of these 324 trends, 184 were political (57%), 106 were commercial (33%), 30 were informational (9%) and four were personal (1%). Of political trends, 116 explicitly concerned the Black Lives Matter protests and catalysing events and 24 concerned COVID. We conducted intercoder reliability checks on the 100 most popular of these trends. Percentage agreement for trend topic was 78% with a Kappa of 63.46%. Percentage agreement for being about BLM was 87.21% with a Kappa of 74.22%. Percentage agreement for being about COVID was 89.95% with a Kappa of 42.96%. The Kappa for being about COVID is lower due to the fact that COVID trends were infrequent in the dataset (7%), meaning that expected agreement for this code is very high (81.19%).

The relative paucity of COVID discourse in the dataset is also notable for this research project. We can compare the focus of trending topics during this weeklong period to two periods analysed in previous research using the same dataset and methods. During the period 22 through 26 April, 13% of the top 200 trends on US Twitter concerned COVID and during the period 27 April through 3 May 19% of the top 200 trends concerned COVID (Bolsover, 2020; Bolsover & Tokitsu Tizon, 2020). However, during the week-long period after the death of George Floyd only 10% of the top 200 trends concerned COVID (almost half that of the week 27 April through 3 May).

During the week of 25 to 31 May, hashtags and keywords associated with the BLM movement made up almost one third of the top 200 trending topics. Only one of these trends (0.5%) was coded as concerning both the BLM movement and COVID. This research set out to understand how protestors balanced and negotiated the twin but linked issues of BLM and COVID, but, these data suggest that Twitter discourse about the BLM movement crowded out or replaced discourse about COVID. However, we need to look in detail at the content of individual posts in order to understand this in more depth.

**Nature of political posts during the period**

We examined a random selection of 250 tweets from within top political trending topics in the week-long period after Floyd's death. We consider the content of all 184 trends that concerned political issues from within the 324 that were trending more than 50 times during the week rather than focus on COVID or BLM trends specially. This is because we are interested in the potential intersection of political, COVID and BLM discourse and the place of BLM discourse within a wider system.

Of these 250 posts, 187 (74.8%) were about politics in the US and 148 (59.2%) were about the BLM movement. Only 13 of the 250 posts were about the coronavirus in the US (5.2%). This is dramatically lower than the period 22 to 26 April in which 35% of 250 randomly selected tweets in the top 200 trends about politics concerned COVID in the US (Bolsover & Tokitsu Tizon, 2020). This, again, suggests that rather than discussing protest action occurring within the context of a global pandemic, discussion of the BLM protests replaced discussion of the coronavirus, despite the ongoing threat of the disease and the potential for discussing the similar root causes of risk that has meant African American communities have disproportionally suffered both from police violence and COVID.

Of the 250 tweets, only one was coded as concerning both COVID in the US and the BLM protests and it only did so in a cursory way, writing *Raw Footage but Necessary* 🖤✊🏾 *#BlackLivesMater #JusticeForGeorgeFloyd #minneapolisriots #AllLivesMatters #coronavirus* with an attached video showing various acts of police brutality against African Americans and footage from the protests, with audio of a speech by Trump. Although using the coronavirus hashtag, the content of the tweet did not concern the disease and it is therefore likely that this hashtag was used to raise the prominence of the tweet rather than to indicate COVID-topical content.

This tweet also contains two discursive elements worthy of note. Firstly, here we see the AllLivesMatter hashtag being used alongside BlackLivesMatter and JusticeforGeorgeFlyod hashtags. This contrasts with the use of the hashtag discussed in research by Carney (2016) in which AllLivesMatter was largely used by those opposing the BLM movement to suggest that BLM was devaluing other lives and to mask racist claims with colour-blind rhetoric. This is clearly not true for this tweet and, perhaps, suggests that this hashtag has been co-opted to support rather than oppose BLM aims.

A second interesting point is the use of the MinneapolisRiots hashtag. As discussed in the previous memo in this series (Bolsover, 2020), labelling is often used on Twitter to assign polarised identity positions to a particular issue. However, we more frequently see the word riot applied to political activity by those who oppose that activity than by those who support that activity (who would more frequently use terms like protest). This, again, suggests a co-opting of oppositional language by supporters of the BLM movement and also, perhaps, support for and admission of an escalation in movement tactics towards violent and unlawful (riot) rather than peaceful and lawful (protest). The experiences of previous BLM protests, such as those in Ferguson after the death of Michael Brown, may have hardened the response of BLM protestors towards violent and uncontrollable protest action.

Returning to the main focus of this research, the fact that this was the only post from 250 to mention, even in passing, both the BLM protests and COVID is notable. This, again, indicates that evidence of an attempt to balance the risks of disease spread against the risks of systemic racism and police violence were not present in BLM discourse on social media. However, this risk balancing might not be overly stated but rather evidenced in the orientation of posts. If a post supported the BLM movement but not protest action, it could be seen as an attempt to find alternative avenues for protests in the time of a global pandemic.

Of the 148 posts concerning the BLM movement, 93 (63%) supported the movement, 17 (12%) opposed the movement and 38 neither supported nor opposed the movement (26%). This is consistent with previous research that found that there was wide support for the movement on social media (Freelon et al., 2016) but is also notable that this support continued during a period where the protest activities of the movement would have increased risk of virus transmission.

Of the 93 posts that supported the BLM movement, 70 (75.3%) explicitly supported protest action. None of these mentioned modifying protest action to restrict the spread of COVID. Of the 51 posts that opposed protest action by the BLM movement, none mentioned COVID as a reason. The hypothesis put forward that protestors would balance risks of disease transmission and systemic racism, and consider the intersectionality of these risks finds no support based on the evidence in this dataset. Rather support or opposition to protest action aligned with support or opposition to the BLM movement more generally. A number of BLM supportive posts contained elements of opposition to protest action, however, this opposition focused entirely on the levels of violence at the protests not on disease risk.

For instance, one tweet showed an Afro-Latino CNN reporter being arrested by police on air, another showed a video of police aggressively approaching the protest line and a third from LA Times reporter Brittny Mejia stated *"I've covered protests involving police in Ferguson, Mo., Baton Rouge, La., Dallas and Los Angeles. I've also covered the U.S. military in war zones, including Iraq and Afghanistan. I have never been fired at by police until tonight."*

Several posts mentioned the need to stay safe, but this was always about staying safe from the police not from an infectious respiratory disease: *"Finally home from the Miami protests in Bayside. We were thousands strong and peaceful as fuck and then shit got rowdy a bit after I left and people are being tear gassed, rubber bullets are flying from snipers and police are in FULL gear.*

Another post shared a photo of what look like two men in battle fatigues with a sniper rifle and other military equipment on the roof of a building, saying: *this is on the corner of 72nd and dodge at omaha. please stay safe*

Throughout the posts, disease risk was not present in discussions with the focus on the risks of violence (largely from the police) at protests. This suggests implicitly that police violence against protestors was seen as the greatest (and perhaps only) risk, with police violence against individual African Americans (against which the protests were mobilised) the second greatest risk. Posters either did not believe the risks of COVID were relevant compared to these other risks, or largely did not think of COVID risk as the focus of media and popular discourse shifted to the BLM movement, or, perhaps, did not wish to discuss or be seen to consider the risks of the disease for fear of detracting or being seen to underplay the risks of police violence and systemic racism towards African Americans in the US.

**Common framings within BLM discourse**

Of the 93 posts that supported the BLM movement, by far the largest proportion (42 posts, 45%) expressed solidarity with the movement's aims and messages. Eleven posts (12%) discussed protest and movement tactics and 10 (11%) criticised police violence and militarisation. A further four posts concerned police violence and militarisation but praised the actions of particular police departments in support of protestors as examples of desirable and possible alternatives. These categories mirror those found in a study of BLM Twitter discourse in 2014 (Ince et al., 2017).

However, there were a further two significant categories of posts that were not prominent in this work. The first category specifically called out white racism and white racist reactions to the BLM movement and black people more generally. Of these nine posts (10%), a number specifically concerned another incident on 25 May in which a white woman in New York called the police after a black birdwatcher approached her to attempt to get her to leash her dog in an area where leashing is required. Others took a more general tone, such as sharing old footage of white children driving black children out of "their" neighbourhood or contemporary footage of a white man in a private gym calling building security because he believed the black men using the gym did not have the right to be there. These posts were considered different from the solidarity posts category as they focus on white racism more generally rather than the specific issue of structural racism in the law enforcement and judiciary towards which the BLM movement is orientated.

A second category of posts also not found in the 2014 research specifically framed their pro-BLM discourse around anti-Trump positions. Both of these categories speak to the opposition and challenges the BLM movement has faced.

Of the 17 posts that opposed the BLM movement protests, the most prominent category was conspiracy content. Interestingly, this evidenced both pro- and anti-BLM sentiments. On the anti-BLM side, two posts argued that the protests had been organised by George Soros, a common scapegoat of far-right voices. Another post wrote: *Minnesota State Police arresting the CNN clown was 100% staged This entire thing is staged by the Democrats to cause chaos and division in our Country Dems won't stop until they've burned our Country to the ground #MinneapolisRiot #minneapolisriots*

In this post, we see much of the same information being repeated in anti-BLM commentary on Twitter as was found in Trump-supporting and anti-lockdown discourse in previous data memos (Bolsover, 2020; Bolsover & Tokitsu Tizon, 2020). George Soros, the Democrats and the traditional media are seen as the enemies and these posts appear to see each new event as a renewed chance to find evidence for this narrative and repeat familiar accusations of the parties behind societal events. This kind of discourse is typical of authoritarian populism that is based on strict in- and out-group constructions and a simplified and visible out-group that threatens the safety of the in-group.

Interestingly, conspiracy-style content was also shared by pro-BLM voices, who alleged that the violence seen at the BLM protest was orchestrated by white people, possibly as a way to discredit the protestors and instigate more widespread violence. Two posters put forward the idea that far-right extremists were attempting to turn the BLM protests into a "New Civil War." Two other posts recounted stories of violence against police cars being instigated by white individuals. One of these posts alleged that these individuals were "plants" trying to bait the crowd into violence and disorder.

This perspective dovetails with the wider commentary on police violence and militarisation, protest tactics and white racism. It evidences a tension among BLM supporting posts between a desire to maintain or evidence a peaceful protest and the feeling of a need for - and even a joy in - lawlessness and violence. Many individuals shared pictures of large groups of orderly and peaceful protesters with captions such as "this needs to be just as viral as the Target video" (of individuals looting a Target store). Another similar post, with an attached picture, wrote:

*POWERFUL! Don't let the news media only show one side of this. A group of black men protecting a police officer who was separated from his group. #riots2020 #protests #JusticeForGeorgeFlyod*

Implicit in these framings are two key and linked narratives which run through wider BLM discourse in the dataset. The first relates to a critique of news media that they will focus largely on the violence of protestors (rather than peaceful and lawful protest) and that they will not show acts of police brutality against protestors. Typifying this perspective, one poster wrote:

*they don't show any of this on the news. spread it all #LAProtests #NashvilleProtest #protest #BlacklivesMaters #pittsburghprotest #phillyprotest #seattleprotest #georgesfloyd #NewarkProtest #buffaloprotest*

The post shared a compilation video of 14 clips of police violence and apparently excessive use of force against protestors, including several scenes of driving into protestors with police cars, trampling a protestor with a police horse and a number of apparently unprovoked physical attacks against protestors. This again demonstrates the importance of cell phone video and social media distribution as documentation of the events that have driven the BLM movement.

A second implicit narrative in these posts is a comparison of police violence against protestor violence. Many posts seem to acknowledge that protestors have acted violently, some even openly embracing the idea of rioting. However, this is held up against documented cases of police violence as a legitimate and necessary response to current conditions. Other posts, such as the aforementioned post of black men protecting a lone white police officer, juxtaposes the non-violent actions of BLM protestors against the violent and militarised police response.

However, it is also important to remember here that a number of posts mentioned cases of police solidarity with protestors – in Camden, New Jersey and Flint, Michigan. These posts demonstrate a path towards a solution for the underlying issues of the movement in which police and activists work together to address issues of systemic racism.

**New issue, same polarised positions**

This data memo set out to investigate, using social media data, how individuals balanced the risks of systemic racism with disease transmission during the major protests that swept the US following the death of George Floyd at a time in which the country was trying to control the spread of COVID-19.

Considering both the spread of trending topics and a random selection of posts in political trends, we find almost no discussion of COVID. The hypothesised balancing of risks is present neither explicitly nor implicitly in either pro- or anti-BLM discourse. The volume of discourse about COVID in trending topics and within political trends dropped from previously examined periods, suggesting discussion of the BLM protests crowded out or replaced discussion of COVID.

However, while the content of these posts does not intersect topically, they evidence the same polarised lenses and the same common enemies of previously analysed COVID discourse. The same arguments that were present in (right-wing) anti-lockdown discourse – alleging conspiracy by the Democrat Party and donors like George Soros, and holding up small business owners as the real victims – were transferred to opposition to the BLM protests.

Similarly, practices of documentation and witness and holding up these behaviours as patently unjustifiable in BLM posts without engaging discussion about why this is the case, were also common practices in previously analysed (left-wing) pro-COVID restriction discourse. One new aspect of BLM case, however, was a shared distrust of traditional media and shared spreading of conspiracy-style allegations on both sides, compared to previous cases in which this content was much more prevalent in right-wing perspectives.

Like our two previous data memos considering reactions to health misinformation about COVID and pro- and anti-COVID restriction discourse, both sides also evidenced worrying levels of divisiveness and hate. Of the 93 posts that supported the BLM movement, 37 (40%) saw distinct opposing groups in society (other than those of political parties). Of these, 31 (84%) appeared to be directing hate at the opposing group. Of the 17 posts that opposed the BLM movement, 8 (47%) saw distinct opposing groups in society and all of these (100%) directed hate towards these groups. Similarly, of the 93 pro-BLM posts 38 (41%) attempted to prevent others from speaking or undermine the value of their words. Thirteen of the 17 anti-BLM posts (76%) attempted to prevent others from speaking or undermine the value of their words. These levels of polarisation, divisiveness, hate and incivility mean that discussion, interaction and productive discourse would be unlikely.

**Concluding remarks and further directions**

This short data memo set out to investigate online discourse about the May 2020 Black Lives Matter protests in the US and to understand how individuals balanced risks of disease transmission against risks of systemic racism in discussing these protests. It found no evidence of this kind of risk balancing occurring or evidence of engagement with the intersectionality of these two risks for African American and ethnic minorities in the US.

Instead, discussion of the BLM protests appeared to replace COVID discussion on the platform, with the number of posts about COVID in political trending topics falling from 35% the week of 27 April to 3 May to 5% in the week following the death of George Floyd. During this period, 59% of posts in political trends concerned the BLM movement. This could have occurred for many reasons: individuals on both sides could see issues of systemic racism and race relations in the US as much more important than COVID. They could also simply be bored of COVID discussion and eager to engage in this new political issue. This also mirrors traditional media coverage, which had previously been dominated by COVID, but turned its focus to the BLM protest during this period. It could also have been the case that the risks of COVID transmission were also on the minds of those participating in and discussing the BLM movement on social media but that they did not want to detract attention away from the issues of system racism and police violence by considering COVID risks and the intersectionality of these risks for minority populations.

However, while the topical focus turned to the BLM protests, replacing discussion of COVID on Twitter, the content of these discussions showed many similarities to COVID discourse in the weeks prior. Discourse on both sides evidenced high levels of polarisation, divisiveness, incivility and hate.

Right-wing political voices blamed a conspiracy by the democratic party, traditional media and democratic donors such as George Soros for the ongoing protests in a very similar way to how right-wing voices understood COVID, COVID control restrictions, and reactions to health misinformation spread by Donald Trump. Pro-BLM voices also included some conspiracy-style content as well as evidencing similar positioning practices as in pro-lockdown discourse (with an assumption rather than argumentation of the rightness of the position).

Compared to studies of BLM discourse from earlier years, pro-BLM discourse during the May 2020 period appeared to have gained new concerns associated with the opposition to the movement from some white Americans, President Trump, and police militarisation and responses to the BLM protests.

In this, we see how structurally as well as discursively the 2020 BLM protests saw existing divisions and polarised positions repeated in a new event. This entrenched positioning, in which each new event is simply a replication of existing divisions in society which predictably align in the new event and play out according to an already familiar story, this is extremely detrimental to the possibility of political progress and potential solution to ongoing political issues.

It is understandable that African Americans, having lived their lives under conditions of systemic racism, would feel a sense of forgone conclusion in these events. However, the posts holding up examples of solidarity and de-escalation tactics from American police departments show some hope for solution in this area. With BLM discourse on Twitter inciting a similar playbook of accusations of Democratic, media and socialist conspiracy from BLM opposers, social media likely will not be a place where issues raised by the BLM movement can be addressed. The platform will likely continue to remain important in distributing documentary evidence of cases of racism and excessive use of force by police, which in turn maintain high levels of support for the movement among the general population. However, especially under the divisive rhetoric of President Trump that has empowered far-right voices, the platform is not one in which discussions seem to be possible that would help solve issues of racism (structural or otherwise) in the general population. Neither was it a place where the complicated balance or risks between systemic racism and disease transmission could be addressed.

Based on this finding, we suggest that it will be difficult to solve either the issues associated with COVID or those highlighted by the BLM movement until the underlying polarisation of these discussions is addressed. Solutions to these entrenched positions, through which new events are perceived as opportunities for the same polarised rhetoric, are essential in addressing the events through which this polarisation is channelled.

It is important to remember, however, that this data memo has focused on a small random sample of social media posts, from a single weeklong period on a single social media platform. It is not intended to provide the final word on these complex issues but rather to form a basis for discussion and further investigation. The US faces many serious and conflicting structural issues, of which only two were addressed here. It is hoped that research such as this that can illuminate how individual Americans are discussing these complicated issues can provide some evidential basis to construct equitable solutions to these ongoing challenges.

## Acknowledgements

I wish to thank Emma Briones, Rhian Hughes and Janet Tokitsu Tizon for their content analysis work on these data; the University of Leeds Strategic Research Investment Fund for helping pay for Emma Briones' and Rhian Hughes' work on the project; and the Laidlaw Scholarship program for funding and facilitating Janet Tokitsu Tizon's work on the project. I also wish to thank Google for providing Cloud Computing research credits that allowed data collection and processing to be undertaken for this research at a time when research funding has been severely curtailed due to COVID-related financial uncertainty.